\documentclass[twocolumn,prb,showpacs,preprintnumbers,amsmath,amssymb]{revtex4}

\usepackage{graphicx}
\usepackage{dcolumn}
\usepackage{bm}

\begin{document}

\preprint{}

\title{Quasiparticle Structure in Antiferromagnetism around 
the Vortex and Nuclear Magnetic Relaxation Time}

\author{Mitsuaki Takigawa}
 \email{takigawa@mp.okayama-u.ac.jp}
 \homepage{http://mp.okayama-u.ac.jp/~takigawa/}
\author{Masanori Ichioka}%
\author{Kazushige Machida}%
\affiliation{
Department of Physics, Okayama University, Okayama 700-8530, Japan}%

\date{\today}

\begin{abstract}
On the basis of the Bogoliubov-de Gennes theory 
for the two-dimensional extended Hubbard model, 
the vortex structure in $d$-wave superconductors 
is investigated including the contribution of the induced 
incommensurate antiferromagnetism around the vortex core. 
As the on-site repulsive interaction $U$ increases, 
the spatial structure of charge and spin changes 
from the antiferromagnetic state with checkerboard modulation to that 
with the stripe modulation. 
By the effect of the induced antiferromagnetic moment, 
the zero-energy density of states is suppressed, and 
the vortex core radius increases. 
We also study the effect of the local density of states (LDOS) change on  
the site-dependent nuclear relaxation rate $T_1^{-1}({\bf r})$. 
These results are compared with a variety of 
experiments performed on high $T_c$ cuprates.
\end{abstract}

\pacs{76.60.Pc, 74.25.Qt, 74.25.Jb, 74.25.Ha}
\maketitle

\section{\label{sec:intro}introduction}

Much attention has been focused on vortex states
in various type II superconductors.
We are now realizing that Fermionic excitations, or 
low-lying quasi-particles induced around a vortex core play a 
fundamental role in determining various physical properties 
of a superconductor and reflect sensitively the pairing 
symmetry.\cite{pair,Franz,d-vortex} 
This is particularly true for $d$-wave pairing case realized in 
high $T_c$ cuprates.  
There, the contribution of nodal quasi-particles is present 
in the spectrum of the quasiparticles, even before applying a field. 

Recently, several experiments report an ``anomalous" vortex core 
in high $T_c$ superconductors, which is not expected in an ordinary 
mixed state:
(1) Hoffman, {\it et al.}\cite{hoffman} made an scanning tunneling 
microscopy (STM) observation and found a checkerboard pattern 
around a vortex core in Bi2212, indicating that the low energy LDOS 
is modified with four-site periodicity. 
(2) Elastic neutron scattering observes the enhancement of the scattering 
peaks associated with incommensurate antiferromagnetism (AFM) modulation in 
{\rm La$_{2-x}$Sr$_x$CuO$_4$}(LSCO)\cite{lake,katano,khaykovich} 
under small fields ($\sim$ a few T) compared with $H_{c2}$. 
The magnetic moment is found to be increased by a small field. 
(3) Recent $\mu$SR measurement detects a static moment 
in the vortex state in under-doped 
{\rm YBa$_2$Cu$_3$O$_{7-\delta}$} (YBCO).\cite{sonier} 
Kadono, {\it et al.}\cite{kadono} 
also demonstrate that the spontaneous moment is 
induced by a small field around a vortex core on LSCO.
(4) Finally, site-selective nuclear magnetic resonance (NMR) 
experiments\cite{halperin,kakuyanagi} 
show that nuclear magnetic relaxation time $T_1$ at the core site 
becomes longer as $T$ decreases, indicating the lack of LDOS at low $T$.
The above facts strongly suggest that in high $T_c$ superconductors 
the vortex core is quite different from what we expect 
in ``normal'' $d$-wave vortex,\cite{d-vortex} 
namely a picture that under a small applied 
field the spatially modulated AFM moments are induced centered 
around a vortex core and simultaneously the zero-energy peak (ZEP)\cite{hess} 
at a core is suppressed. 
This picture is also consistent with the earlier STM 
observations which show an ``empty core", namely the absence 
of the ZEP at the core site.\cite{renner} 
The AFM around the vortex 
\cite{arovas,zhang,franzQED3,ogata,kishine,tsuchi,demler,martin,takigawaPRL2003,zhu,chen}
was also studied theoretically 
in SO(5) model,\cite{arovas,zhang} 
QED$_3$ model,\cite{franzQED3} $t$-$J$ model,\cite{ogata,kishine,tsuchi}  
and Hubbard model.\cite{martin,takigawaPRL2003,zhu,chen} 

Quite recently, two new NMR imaging experiments\cite{kakuyanagi2,mitrovic} 
on the $T$-dependence of $T_1$ are reported: 
Kakuyanagi, {\it et al.}\cite{kakuyanagi2} show 
in Tl$_2$Ba$_2$CuO$_6$ that as lowering $T$, 
$1/(T_1T)$ at a core site increases divergingly toward a temperature 
$T_M$ below $T_c$ and then decreases at lower $T$.
This is contrasted with $1/(T_1T)$ at other sites which exhibit a monotonic 
decrease below $T_c$. 
This result suggests that below $T_M$ the local magnetism 
appears exclusively at a vortex core site and other sites stay in 
the normally expected $d$-wave state.
Mitrovi\' c, {\it et al.}\cite{mitrovic} 
observe a similar divergent behavior 
in $1/(T_1T)$ of YBa$_2$Cu$_3$O$_7$ at the core site 
where the crossover temperature $T_M$ is quite low, 
and $1/(T_1T)$ at the outside sites of the core 
show a constant at low $T$. 
The implication of these experiments is twofold: 
The ZEP in $d$-wave vortex core must be present 
at $T_M<T<T_c$ and in some $H$ region, and its contribution enhances 
$1/(T_1T)$ in this temperature region. 
Below $T_M$, the field induced local AFM must exist, 
which is to remove the  ZEP, giving rise 
to a suppression of $1/(T_1T)$ at lower $T$.
These new experiments further enforce the above picture 
of the AFM-induced vortex core state. 

Here we explore a possibility of the induced incommensurate AFM 
moment around a vortex in $d$-wave pairing state.
It is quite possible because 
(A) in a $d$-wave vortex state the zero-energy quasi-particles 
produced at a vortex core are readily available 
for the magnetism and 
(B) these quasi-particles are under strongly correlated 
environment in cuprates, where the AFM state appears near half filling 
next to the superconducting phase. 
Therefore, based on these experimental facts we are led to investigate 
how two orders; $d$-wave superconducting order with vortices and the 
modulated magnetic order can be influenced in order to compromise 
mutual competing effects.

The purposes of this paper are to study detailed properties of 
the vortex core under the possible magnetic order, in particular 
on the quasi-particle structure 
and to calculate the site-selective $T_1({\bf r})$ 
which turns out to be a good probe for examining 
the quasi-particles at a vortex core.
This study is not only 
one of our continuous efforts on the vortex problem 
but also associated with our stripe problem.\cite{machida,oka} 
When the stripe structure is formed, 
the doped carriers are accommodated neatly at the stripe region, 
ultimately relating to the mechanism of the high $T_c$ superconductivity.
Therefore, we consider the case when the induced AFM around vortex 
is not in commensurate AFM structure\cite{zhu,chen} 
with the ordering vector ${\bf Q}=(\pi,\pi)$, 
but in incommensurate spin and charge structure, where 
the spin ordering vector is given by 
${\bf Q}=(2\pi(\frac{1}{2}-\epsilon),\pi)$ or 
${\bf Q}=(\pi,2\pi(\frac{1}{2}-\epsilon))$. 
Following the neutron scattering results,\cite{lake,katano,khaykovich} 
$\epsilon=\frac{1}{8}$ at hole filling $n_{\rm h}\sim \frac{1}{8}$, 
meaning the eight site periodic spin structure.
In this case, the charge structure characterized by the ordering vector 
$2{\bf Q}$ shows four site periodic oscillation, which is consistent with 
the STM observation.\cite{hoffman} 

The arrangement of this paper is as follows: 
After introducing the model Hamiltonian which allows us to describe 
two possible orderings of superconductivity 
and magnetism in an equal footing, 
namely, the so-called extended Hubbard Hamiltonian, 
we set up the Bogoliubov-de Gennes (BdG) equation on a lattice, and a formula 
for the LDOS and $T_{1}({\bf r})$ in Sec. 2. 
The basic vortex properties, including the amplitude of the induced moment, 
the local charge distribution, and the Fourier component 
are studied in Sec. 3. 
We discuss the LDOS in Sec. 4, and 
evaluate the $T$-dependence and 
site-dependence of $T_{1}({\bf r})$ in Sec. 5.
The final section is devoted to conclusion and discussion.
Some of the results are briefly reported 
in Refs. \onlinecite{takigawaPRL2003}
and \onlinecite{takigawaLT23}.

\section{\label{sec:BdG}Bogoliubov-de Gennes theory on extended Hubbard model}

We begin with an extended Hubbard model on a two-dimensional 
square lattice, and introduce the mean fields 
$n_{i,\sigma}= \langle a^\dagger_{i,\sigma} a_{i,\sigma} \rangle$
at the $i$-site, where $\sigma$ is a spin index and $i=(i_x,i_y)$ and 
$\Delta_{{\hat e},i,\sigma}=\langle a_{i,-\sigma} a_{i+\hat{e},\sigma} \rangle$.
We assume a pairing interaction $V$ between nearest-neighbor (NN) sites. 
This type of pairing interaction gives $d$-wave 
superconductivity.\cite{takigawa,takigawa2,morr} 
Thus, the mean-field Hamiltonian under $H$ is given by 
\begin{eqnarray}
{\cal H}&=&
-\sum_{i,j,\sigma} \tilde{t}_{i,j}a^{\dagger}_{i,\sigma} a_{j,\sigma}
+U\sum_{i,\sigma} n_{i,-\sigma}
a^\dagger_{i,\sigma} a_{i,\sigma} 
\nonumber \\ &&
+V\sum_{\hat{e},i,\sigma}
(\Delta^{*}_{\hat{e},i,\sigma} a_{i,-\sigma} a_{i+\hat{e},\sigma}
+\Delta_{\hat{e},i,\sigma} a^\dagger_{i,\sigma} a^\dagger_{i+\hat{e},-\sigma}
)\qquad, 
\label{eq:2.2}
\end{eqnarray}
where $a^{\dagger}_{i,\sigma}$
($a_{i,\sigma}$) is a creation (annihilation) operator, and 
$i+\hat{e}$ represents the NN site ($\hat{e}=\pm\hat{x},\pm\hat{y}$).
The transfer integral is expressed as 
\begin{eqnarray}
&&\tilde{t}_{i,j}=t_{i,j} \exp [ {\rm i}\frac{\pi}{\phi_0}\int_{{\bf
r}_i}^{{\bf r}_j} {\bf A}({\bf r}) \cdot {\rm d}{\bf r} ], 
\label{eq:BdG4}
\end{eqnarray}
with the vector potential 
${\bf A}({\bf r})=\frac{1}{2}{\bf H}\times{\bf r}$ in the symmetric gauge, 
and the flux quantum $\phi_0$.
The external field is introduced as 
the so-called Peierls phase factor.

We assume the following hopping integrals $t_{ij}$: 
For the NN pairs $(i,j)$, $t_{i,j}=t$.
For the next-NN pairs situated on a diagonal position on the square
lattice, $t_{i,j}=t'$. 
For the third-NN pairs, which are situated along the NN bond
direction, $t_{i,j}=t''$.
To reproduce the Fermi surface topology of cuprates, we set $t'=-0.12t$ and 
$t''=0.08t$.~\cite{tohyama} 
We consider mainly the pairing interaction $V=-2.0t$.
The essential results of this paper do not significantly depend on the 
choice of these parameter values. 

In terms of the eigenenergy $E_\alpha$ and the wave functions
$u_\alpha({\bf r}_i)$ and $v_\alpha({\bf r}_i)$ at the $i$-site,
the BdG equation is given by
\begin{eqnarray}
\sum_j
\left( \begin{array}{cc}
K_{\uparrow,i,j} & D_{i,j} \\ D^\dagger_{i,j} & -K^\ast_{\downarrow,i,j}
\end{array} \right)
\left( \begin{array}{c} u_\alpha({\bf r}_j) \\ v_\alpha({\bf r}_j)
\end{array}\right)
=E_\alpha
\left( \begin{array}{c} u_\alpha({\bf r}_i) \\ v_\alpha({\bf r}_i)
\end{array}\right),
\label{eq:BdG1}
\end{eqnarray}
where
$K_{\sigma,i,j}=-\tilde{t}_{i,j} +\delta_{i,j} (U n_{i,-\sigma} -\mu)$ 
with the chemical potential $\mu$, 
$D_{i,j}=V \sum_{\hat{e}}  \Delta_{i,j} \delta_{j,i+\hat{e}} $ 
and $\alpha$ is an index of the eigenstate.\cite{takigawa,takigawa2} 

We study the case of the square vortex lattice where the NN vortex is
located in the direction of $45^\circ$ from the $a$-axis, 
which is known to be a stable vortex configuration.
The unit cell in our calculation is the square area of
{$N_r$}$^2$ sites where two vortices are accommodated.
Then, the magnetic field is given by 
$H = 2\phi_0 / (a{N_r})^2$ with the lattice constant $a$.  
Thus, we denote the  field strength by $N_r$ as $H_{N_{r}}$.
We consider the area of {$N_k$}$^2$ unit cells.
By introducing the quasi-momentum of the magnetic Bloch state,
${\bf k}=(2\pi /aN_rN_k)(l_x,l_y):(l_x,l_y = 1,2,\cdots,N_k)$,
we set
$u_\alpha({\bf r})=\tilde{u}_\alpha({\bf r})
{\rm e}^{{\rm i} {\bf k}\cdot{\bf r}},
v_\alpha({\bf r})=\tilde{v}_\alpha({\bf r})
{\rm e}^{{\rm i} {\bf k}\cdot{\bf r}}$.
Then, the eigenstate of $\alpha$ is labeled by {\bf k} and 
the eigenvalues obtained by solving Eq. (\ref{eq:BdG1}) within a unit cell.
The periodic boundary condition is given by the symmetry
for the translation ${\bf R}=l_x{\bf R}_x^0 + l_y {\bf R}_y^0$, where 
${\bf R}_x^0=(a N_r,0)$ and ${\bf R}_y^0=(0,a N_r)$ are unit vectors 
of the unit cell for our calculation. 
Then, the translational relation is given by 
$\tilde{u}_\alpha({\bf r}+{\bf
R})=\tilde{u}_\alpha({\bf r}) {\rm e}^{i\chi({\bf r},{\bf R})/2}$ 
and 
$\tilde{v}_\alpha({\bf r}+{\bf R})=\tilde{v}_\alpha({\bf r})
{\rm e}^{-i\chi({\bf r},{\bf R})/2}$ 
with 
\begin{equation}
\chi({\bf r},{\bf R})
= -\frac{2\pi}{\phi_0}{\bf A}({\bf R})\cdot{\bf r}
- 2\pi l_x(l_x-l_y) + \frac{2 \pi}{\phi_0}
({\bf H}\times {\bf r}_0)\cdot{\bf R}
\end{equation}
in the symmetric gauge. 
The vortex center is located
at ${\bf r}_0+\frac{1}{4}(3{\bf R}_x^0+{\bf R}_y^0)$.

The self-consistent conditions for the pair potential $\Delta_{i,j}$
and the number density $n_{i,\sigma}$ are given by 
\begin{eqnarray}
&&\Delta_{i,j}
=\langle a_{j,\downarrow} a_{i,\uparrow} \rangle 
=\sum_\alpha  u_\alpha({\bf r}_i) v^\ast_\alpha({\bf r}_j) 
f(E_\alpha),\\
&&n_{i,\uparrow}=\langle a^\dagger_{i,\uparrow} a_{i,\uparrow} \rangle 
=\sum_\alpha |u_\alpha({\bf r}_i)|^2 f(E_\alpha ), \\
&&n_{i,\downarrow}=\langle a^\dagger_{i,\downarrow} a_{i,\downarrow}  \rangle 
=\sum_\alpha |v_\alpha({\bf r}_i)|^2 (1-f(E_\alpha ))
\end{eqnarray}
with the Fermi distribution function $f(E)$. 
The charge density $n({\bf r}_i)=n_{i,\uparrow}+n_{i,\downarrow}$, 
the spin density 
$S_{z}({\bf r}_i)=\frac{1}{2}(n_{i,\uparrow}-n_{i,\downarrow})$
and the staggered magnetization $M({\bf r}_i)=(-1)^{i_x+i_y}S_{z}({\bf r}_i)$. 
The $d$-wave order parameter at site $i$ is

\begin{eqnarray}
&&\Delta_{d}({\bf r}_i)=
( \Delta_{ \hat{x},i}
 +\Delta_{-\hat{x},i}
 -\Delta_{ \hat{y},i}
 -\Delta_{-\hat{y},i} )/4
\end{eqnarray}
with
$\Delta_{\hat{e},i}={\bar \Delta}_{i,i + \hat{e}}
\exp[{\rm i}\frac{\pi}{\phi_0}
\int_{{\bf r}_i}^{({\bf r}_i+{\bf r}_{i + \hat{e}})/2}
{\bf A}({\bf r}) \cdot {\rm d}{\bf r}],
$
%
where the singlet pairing component 
${\bar \Delta}_{i,i+\hat{e}}=\langle a_{i+\hat{e},\downarrow} 
a_{i,\uparrow} \rangle- \langle a_{i+\hat{e},\uparrow} 
a_{i,\downarrow} \rangle$. 
As the induced order parameter, we can define 
the triplet-$d$-wave order parameter at $i$-site, given by 
\begin{eqnarray}
&&\Delta_{d,i}^{\rm triplet}({\bf r}_i)=
( \Delta_{ \hat{x},i}^{\rm triplet}
 +\Delta_{-\hat{x},i}^{\rm triplet}
 -\Delta_{ \hat{y},i}^{\rm triplet}
 -\Delta_{-\hat{y},i}^{\rm triplet} )/4, \qquad 
\end{eqnarray}
where we use the triplet pairing component 
${\bar \Delta}_{i,i+\hat{e}}^{\rm triplet}=\langle a_{i+\hat{e},\downarrow} 
a_{i,\uparrow} \rangle+ \langle a_{i+\hat{e},\uparrow} 
a_{i,\downarrow} \rangle$ 
instead of ${\bar \Delta_{i,i+\hat{e}}}$.

We construct the Green's functions from $E_\alpha$, $u_\alpha({\bf 
r})$, $v_\alpha({\bf r})$ defined as
\begin{eqnarray} &&
\hat{g}(x,x') 
\equiv \left(
 \begin{array}{cc} 
g_{11}(x,x') & g_{12}(x,x') \\ g_{21}(x,x') & g_{22}(x,x') 
 \end{array}  \right) 
\nonumber  \\ &&
\equiv \left( 
\begin{array}{cc}
-\left< T_{\tau}[\hat{\psi}_{\uparrow}(x)
       \hat{\psi}_{\uparrow}^{\dagger}(x')] \right> & 
-\left< T_{\tau}[\hat{\psi}_{\uparrow}(x)
       \hat{\psi}_{\downarrow}(x')] \right> \\ 
-\left<T_{\tau}[\hat{\psi}_{\downarrow}^{\dagger}(x)
       \hat{\psi}_{\uparrow}^{\dagger}(x')] \right> & 
-\left<T_{\tau}[\hat{\psi}_{\downarrow}^{\dagger}(x)
       \hat{\psi}_{\downarrow}(x')]\right> \end{array} 
\right).\qquad 
\label{eq:green}
\end{eqnarray}
with $x\equiv({\bf r},\tau)$.  
After the Fourier transformation of $\tau$ as
\begin{eqnarray}
\hat{g}(x,x') 
= T\sum_{\omega_n} e^{-i\omega_n(\tau-\tau')}
\hat{g}({\bf{r}},{\bf{r'}},\omega_n), 
\end{eqnarray} 
the thermal Green's functions with the 
Fermionic imaginary frequency $\omega_n=2\pi T(n+{1\over2})$
are written as
\begin{eqnarray} 
g_{11}({\bf{r}},{\bf{r'}},\omega_n) &=& \sum_{\alpha}
\frac{u_{\alpha}({\bf{r}})u_{\alpha}^{*}({\bf{r'}})}
{i\omega_n-E_{\alpha}} , 
\label{eq:green2-1}\\ 
g_{12}({\bf{r}},{\bf{r'}},\omega_n) &=& \sum_{\alpha}
\frac{u_{\alpha}({\bf{r}})v_{\alpha}^{*}({\bf{r'}})} 
{i\omega_n-E_{\alpha}} ,\\ 
g_{21}({\bf{r}},{\bf{r'}},\omega_n) &=& \sum_{\alpha}
\frac{v_{\alpha}({\bf{r}})u_{\alpha}^{*}({\bf{r'}})} 
{i\omega_n-E_{\alpha}} ,\\
g_{22}({\bf{r}},{\bf{r'}},\omega_n) &=& \sum_{\alpha}
\frac{v_{\alpha}({\bf{r}})v_{\alpha}^{*}({\bf{r'}})} 
{i\omega_n-E_{\alpha}}. 
\label{eq:green2-4}
\end{eqnarray}

%
The LDOS is evaluated by using the thermal Green's functions as 
\begin{equation}
N_{\uparrow}(E,{\bf{r}})=
-{1\over \pi}{\rm Im}g_{11}({\bf{r}},{\bf{r}},i\omega_n \rightarrow E+i\eta)
\end{equation} 
for the up-spin electron contribution, and 
\begin{equation}
N_{\downarrow}(E,{\bf{r}})
={1\over \pi}{\rm Im}g_{22}({\bf{r}},{\bf{r}},-i\omega_n \rightarrow E+i\eta)
\label{eq:ldosdown}
\end{equation}
for the down-spin electron contribution. 
Then, the total LDOS is given by 
\begin{eqnarray} &&
N(E,{\bf{r}}) 
= N_{\uparrow}(E,{\bf{r}}) + N_{\downarrow}(E,{\bf{r}}) 
\nonumber \\ && 
=\sum_\alpha \{ |u_\alpha({\bf{r}})|^2\delta(E-E_\alpha)
+|v_\alpha({\bf{r}})|^2\delta(E+E_\alpha) \} . \qquad 
\label{eq:LDOS0}
\end{eqnarray} 
When we consider the differential tunnel conductance of STM experiments, 
the $\delta$-functions in eq. (\ref{eq:LDOS0}) 
are replaced by the derivative $-f'(E)$ of the Fermi distribution 
function $f(E)$:
$N(E,{\bf r})
=-\sum_\alpha [|u_\alpha ({\bf r})|^2f'(E_\alpha -E)
+ |v_\alpha ({\bf r})|^2f'(E_\alpha +E)].$

By calculating the spin-spin correlation
function $\chi_{+,-}({\bf r},{\bf r}',i \Omega_n)$ 
from the Green's functions in Eqs. (\ref{eq:green2-1})-(\ref{eq:green2-4}),
we obtain the nuclear spin relaxation rate,\cite{takigawa2} 
\begin{eqnarray}
R({\bf r},{\bf r}') &=&
{\rm Im}\chi_{+,-}({\bf r},{\bf r}',
i \Omega_n \rightarrow \Omega + {\rm i}\eta)/(\Omega/T)|_{\Omega
\rightarrow 0}
\nonumber \\
&=&
 -\sum_{\alpha,\alpha'} 
[
 u_{\alpha}({\bf r})u^\ast_{\alpha}({\bf r}')
 v_{\alpha'}({\bf r})v^\ast_{\alpha'}({\bf r}')
\nonumber \\ &&
 -v_{\alpha}({\bf r})u^\ast_{\alpha}({\bf r}')
 u_{\alpha'}({\bf r})v^\ast_{\alpha'}({\bf r}')
 ]
\nonumber \\ &&
\times \pi T f'(E_\alpha) \delta(E_\alpha - E_{\alpha'}).
\label{eq:T1}
\end{eqnarray}
We consider the case ${\bf r}={\bf r}'$ by assuming that
the nuclear relaxation occurs at a local site.
Then, ${\bf r}$-dependent relaxation time is given by $T_1({\bf
r})=1/R({\bf r},{\bf r})$.
We use $\delta(x)=\pi^{-1} {\rm Im}(x-{\rm
i}\eta)^{-1}$ to handle the discrete energy level of the finite
size calculation.
We typically use $\eta=0.01t$. In Eq. (\ref{eq:T1}), 
the first term is proportional to 
$N_{\uparrow}({\bf r},E) \times N_{\downarrow}({\bf r},E)$ 
when ${\bf r}={\bf r}'$. 

We typically consider the case of a unit cell with $24 \times 24$ sites, 
where two vortices are accommodated.
The vortex cores are located at 
($i_x$,$i_y$)=(1,1) and ($N_r/2+1$,$N_r/2+1$).
The spatially averaged hole density is set to 
$n_{\rm h}=1-\overline{n_i} \sim \frac{1}{8}$ by tuning the 
chemical potential $\mu$. 
By introducing the quasimomentum of the magnetic Bloch state,
we obtain the wave function under the periodic boundary condition 
whose region covers many unit cells. 
In this paper, temperature $T$ is scaled by $T_c$ 
which is the superconducting critical temperature when $U/t=0$. 

For the initial state of our calculation, 
we give a lowest Landau level function of the vortex state for $\Delta$, 
and very weak incommensurate spin modulation for $n_\sigma$ with the ordering 
vector ${\bf Q}=(\pi,2\pi(\frac{1}{2}-\epsilon))$ and $\epsilon =\frac{1}{8}$ 
appropriate to the hole filling $\frac{1}{8}$. 
By iterating the calculation, $M({\bf r}_i) \rightarrow 0$ 
when the AFM is absent. 
When the AFM appears, $M({\bf r}_i)$ grows and makes checkerboard or 
stripe pattern depending on the parameters. 

\section{\label{sec:Vortex}Properties of antiferromagnetism around the vortex core}

We solve the BdG equation self-consistently 
with the on-site repulsive interaction $U$, 
using an eight-site periodic vertical stripe state as an initial state.
We consider the case that spin and charge modulations appear around the 
vortex core, which is stabilized by $U$.
Then the self-consistent solution of the AFM-induced $d$-wave vortex is 
analyzed and compared with the ``normal'' $d$-wave vortex for $U=0$.
\cite{takigawa,takigawa2}

	\subsection{\label
	{sec:subsec1}Induced moment around the core}

\begin{figure*}[thb]
\includegraphics[width=16cm]{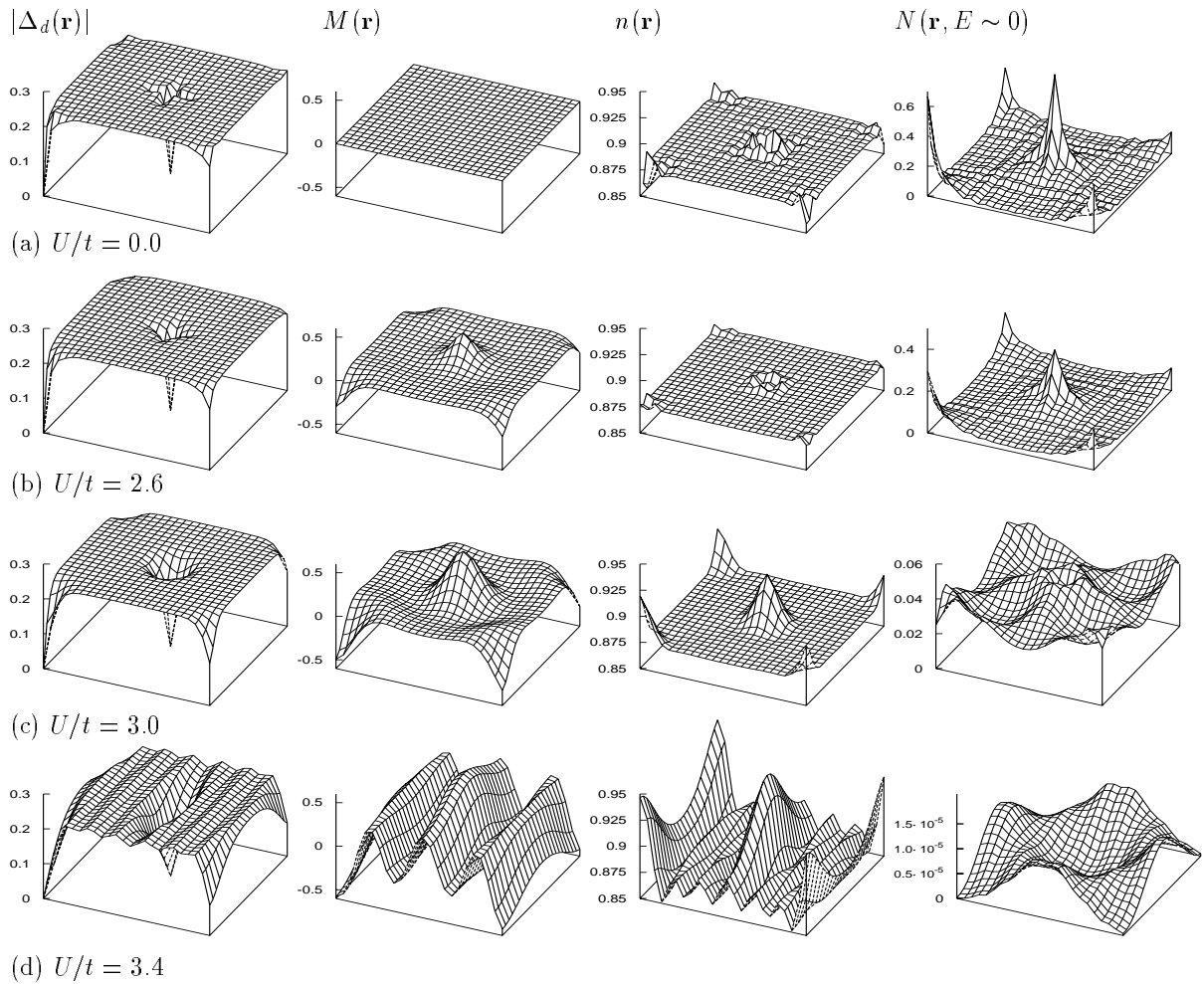}
\caption{\label{fig:r}
Spatial variation of the vortex state at $T=T_c/40$ 
for $U/t=$0.0 (a), 2.6 (b), 3.0 (c) and 3.4 (d). 
From left to right, we show 
the amplitude of the $d$-wave oder parameter $|\Delta_d({\bf r}_i)|$, 
the staggered magnetization $M({\bf r}_i)$, 
the charge density $n({\bf r}_i)$, 
and the zero-energy LDOS $N(E=0,{\bf r}_i)$ within a unit cell. 
The vortex center is located at the center $(i_x,i_y)=(N_r/2+1,Nr/2+1)$ 
and a corner(1,1).
}
\end{figure*}

Figure \ref{fig:r} shows the spatial structure of vortex state, 
i.e., the amplitude of the $d$-wave oder parameter $|\Delta_d({\bf r}_i)|$, 
the staggered magnetization $M({\bf r}_i)$, 
the charge density $n({\bf r}_i)$, 
the zero-energy LDOS $N(E=0,{\bf r}_i)$, 
within a unit cell for various $U$'s.

First, we consider the induced magnetic moment around the vortex core. 
In $d$-wave vortex case without AFM at $U=0$,  
the superconductivity $|\Delta_d({\bf r})|$ is suppressed at the vortex core 
and the zero-energy LDOS has peak there (Fig. \ref{fig:r}(a)). 
Using these low energy states, the staggered moment 
is induced centered at the vortex core for larger $U$ case. 
It is seen from Fig. \ref{fig:r} that 
(1) At $U=0$ no moment appears. 
(2) The induced moment appears centered around 
the core at $U > U_{\rm cr}$, and the moment grows as $U$ increases.
(3) The spatial structure of the induced moment changes 
from a simple AFM type to a stripe type when $U>3.1t$. 
In the former simple AFM state has also checkerboard type modulation. 
This checkerboard modulation is barely seen in Fig. \ref{fig:r}(c), 
where the ``floor'' in $M({\bf r})$ is modulated with eight-site period.
(4) In the stripe case (Fig. \ref{fig:r}(d)) 
the amplitude maximum of $M({\bf r})$ coincides with the vortex core in this example.
These tendencies are already seen in our previous paper 
for the vortex state in the stripe state.\cite{oka} 

\begin{figure}[htb]
\includegraphics[width=8cm]{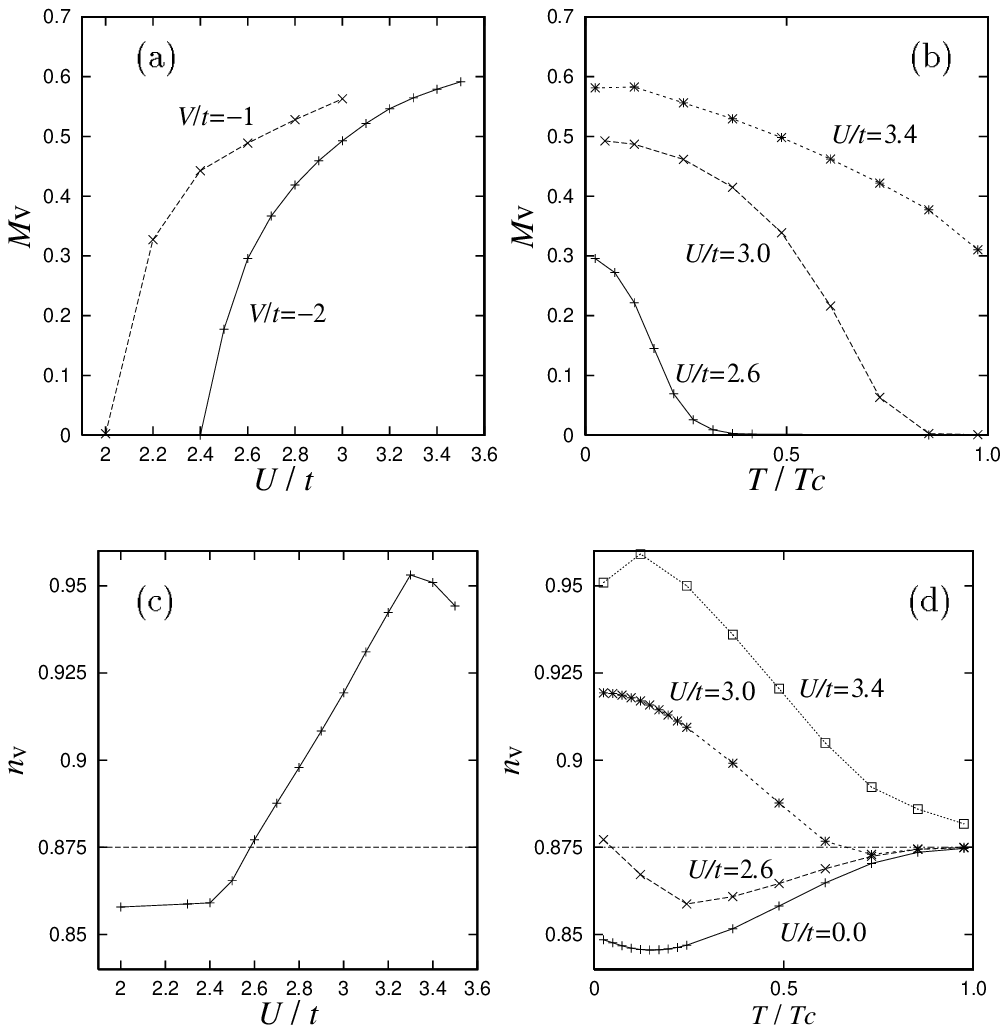}
\caption{\label{fig:UT}
The AFM moment $M_{\rm v}$ and the charge density $n_{\rm v}$ 
at the vortex center. 
(a) $U$-dependence of $M_{\rm v}$ at $T=T_c/40$ for $V/t=-1$ and $-2$. 
(b) $T$-dependence of $M_{\rm v}$ for $U/t=0$, 2.6, 3.0 and 3.4. 
(c) $U$-dependence of $n_{\rm v}$ at $T=T_c/40$ for $V/t=-2$. 
(d) $T$-dependence of $n_{\rm v}$ for $U/t=0$, 2.6, 3.0 and 3.4.  
}
\end{figure}

In Fig. \ref{fig:UT}(a), we show the AFM moment at the core  $M_{\rm v}$ as 
a function of $U$ for the two cases $V/t=-1$ and $V/t=-2$.
The induced moment appears above the critical $U_{\rm cr}$, 
which depends on $V$.
The critical strength $U_{\rm cr}$ is larger for stronger pairing 
interaction $V$, as is seen from Fig. \ref{fig:UT}(a),  
because superconductivity tends to suppress the AFM moment.
When $U < U_{\rm cr}$, $M_{\rm v}=0$ and 
the vortex structure is the same as $U=0$ case.
We show the $T$-dependence of $M_{\rm v}$ for several $U$ values 
in Fig. \ref{fig:UT}(b). 
For $U/t=3.4$ the moment appears above $T_c$. 
In this strong $U$ case, the spatial structure of the 
moment is a stripe type (see Fig. \ref{fig:r}(d)), 
and at $T>0.5T_c$, the stripe periodicity changes from eight site period 
by the temperature effect. 
For smaller $U$, 
$M({\bf r})$ becomes a checkerboard type as shown in Fig. \ref{fig:r}(c), 
and $M_{\rm v}$ appears at $T_M$ below $T_c$. 
With decreasing $U$, $T_M$ decreases.

	\subsection{\label
	{sec:subsec3}Charge density around the core}

The spatial variation of the charge density $n({\bf r})$ around 
the vortex core is shown also in Fig. \ref{fig:r}, 
which is also affected by the induced AFM. 
While $n({\bf r})$ is suppressed 
at the vortex core at $U=0$,\cite{hayashi2} 
$n({\bf r})$ increases with approaching the core, 
when $M_{\rm v}$ appears at $U>U_{\rm cr}$.
These properties of the charge structure are related to the modulation 
of the LDOS, as discussed later. 
In the large $U$ case giving $T_M>T_c$, 
the stripe structure becomes eminent and the 
charge stripe structure is seen in Fig. \ref{fig:r}(d).  
When comparing with the spin stripe structure of $M({\bf r})$, 
$n({\bf r})$ has minimum at the node of the $M({\bf r})$ oscillation, 
implying that the excess hole carriers accumulate there.
The periodicity of the charge modulation is four-site, 
in our doping level $n_h \sim 1/8$. 
It is half of the spin structure periodicity.

The $U$-dependence of the charge at the core $n_{\rm v}$ is 
shown in Fig. \ref{fig:UT}(c), 
where we see that the suppressed $n_{\rm v}$ at $U<U_{\rm cr}$ 
is increased with the growth of $M_{\rm v}$ when $U>U_{\rm cr}$. 
The $T$-dependence of $n_{\rm v}$ is shown in Fig. \ref{fig:UT}(d).
When $U=0$, $n_{\rm v}$ is reduced around the vortex core 
at low temperature because of the particle-hole asymmetry.\cite{hayashi2} 
For $U>U_{\rm cr}$, the decreasing dependence of $n_{\rm v}$ 
on lowering temperature changes to the increasing one at $T<T_M$ 
where the moment begins to appear.
These charge behavior is similar to the case when commensurate AFM 
is induced around the vortex core.\cite{zhu,chen}

	\subsection{\label
	{sec:subsec4}Fourier component of the spin and charge modulation}

\begin{figure}[htb]
\includegraphics[width=7cm]{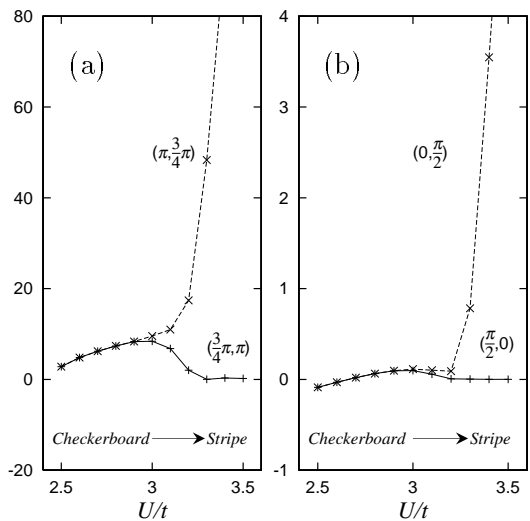}
\caption{\label{fig:FT}
$U$-dependence of the Fourier component. 
(a) Eight-site periodic Fourier component 
of the magnetization, $S_{z,{\bf Q}}$ for ${\bf Q}=(\frac{3}{4}\pi,\pi)$ 
and ${\bf Q}=(\pi,\frac{3}{4}\pi)$. 
(b) Four-site periodic  Fourier component 
of the charge density, $n_{2{\bf Q}}$ for 
$2{\bf Q}=(\frac{1}{2}\pi,0)$ 
and $2{\bf Q}=(0,\frac{1}{2}\pi)$. $T=T_c/40$. 
}
\end{figure}

In order to see the checkerboard modulation of $M({\bf r})$ 
superimposed in the basic AFM variation, 
we calculate the Fourier transformation $S_{z,{\bf q}}$ and 
$n_{\bf q}$ from $S_{z}({\bf r}_i)$ and $n({\bf r}_i)$, respectively.  
In addition to the peak of the vortex lattice period, 
$S_{z,{\bf q}}$ has the peaks at ${\bf q}={\bf Q}_1=(\frac{3}{4}\pi,\pi)$ 
and/or ${\bf q}={\bf Q}_2=(\pi,\frac{3}{4}\pi)$, 
where we set the lattice constant to be unity. 
These ${\bf Q}_1$ and ${\bf Q}_2$ are, respectively, ordering vectors of the 
eight-site period spin structure along $x$ and $y$ directions. 
These peaks correspond to the neutron scattering peaks observed 
in LSCO.\cite{lake,katano,khaykovich} 
As for the charge structure, ${\bf q}$ has the peaks at 
${\bf q}=2{\bf Q}_1$ and/or ${\bf q}=2{\bf Q}_2$. 
In Fig. \ref{fig:FT}(a), we show the $U$-dependence of the peak height 
for $S_{z,{\bf q}}$ at ${\bf q}={\bf Q}_1$ and ${\bf q}={\bf Q}_2$. 
When $U<3.1t$, the peak heights at ${\bf Q}_1$ and at ${\bf Q}_2$ are 
the same, meaning the checkerboard modulation both $x$ and $y$ directions. 
Since the peak height increases, the spin modulation becomes eminent 
with increasing $U$. 
When $U$ becomes larger than $3.1t$, 
one of the peaks for  $S_{z,{\bf Q}}$ vanishes, and 
one of the peaks remains at ${\bf Q}_1$ or ${\bf Q}_2$, 
which means a one-dimensional stripe, 
as is seen in Fig. \ref{fig:r}(d). 

This checkerboard-stripe transition is also seen in the Fourier transformation 
of the charge structure shown in Fig. \ref{fig:FT}(b). 
At $U<3.1t$, $n_{\bf q}$ has the peaks 
both at $2{\bf Q}_1$ and $2{\bf Q}_2$, 
meaning four site periodic checkerboard charge structure. 
At $U \ge 3.1t$, $n_{\bf q}$ has the peak either at $2{\bf Q}_1$ or 
at $2{\bf Q}_2$, meaning four site periodic stripe structure.   

	\subsection{\label
	{sec:subsec5}Vortex core radius}

Here, we discuss the vortex core shape and radius from the 
spatial structure of $|\Delta_d({\bf r})|$. 
As seen from $|\Delta_d({\bf r})|$ in Fig. \ref{fig:r}(a), 
when the AFM is absent at $U < U_{\rm cr}$, 
we see the fourfold symmetric vortex core shape, 
reflecting the fourfold symmetry of the $d_{x^2-y^2}$-wave 
pairing potential in the momentum space.\cite{pair} 
After the AFM appears at $U > U_{\rm cr}$, 
the vortex core becomes circular shape, as shown in Fig. \ref{fig:r}(c). 
In the stripe case at $U/t=3.4$,  $|\Delta_d({\bf r})|$ also shows 
four-site periodic structure, where $|\Delta_d({\bf r})|$ is enhanced 
at the hole-rich site. 
In this case, vortex core shape is enlarged along the stripe direction.

\begin{figure*}[htb]
\includegraphics[width=15cm]{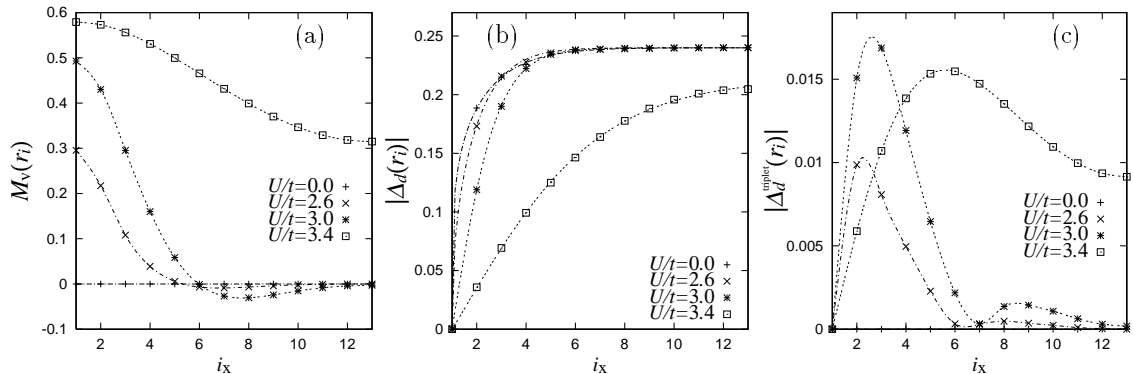}
\caption{\label{fig:OP}
The profile of 
the staggered magnetization $M({\bf r})$ (a), 
the $d$-wave singlet order parameter amplitude 
$|\Delta_{d}({\bf r}_i)|$ (b), 
and the $d$-wave triplet order parameter amplitude 
$|\Delta_{d}^{\rm triplet}({\bf r}_i)|$(c) 
along the $x$ axis direction, $i_y=1$. 
The vortex core is located at $i_x=1$.
$T=T_c/40$. 
$U/t=0$, 2.6, 3.0 and 3.4.
}
\end{figure*}

To discuss the core radius, in Fig. \ref{fig:OP} 
we show the profile of the order parameter $|\Delta_{d}({\bf r}_i)|$ and 
staggered magnetization $M({\bf r}_i)$ along the $x$-direction, 
which is the next NN vortex direction and also the parallel direction to the stripe. 
When the magnetization appears around the core, 
the $d$-wave order parameter is suppressed, 
therefore the radius of the vortex core becomes large.
To estimate the core radius, 
we fit the profile of $|\Delta_{d}({\bf r}_i)|$ 
along the next NN vortex direction by $\Delta_M \tanh^A(r/B)$,  
where $\Delta_M$ is the amplitude in the bulk, 
and $A$ and $B$ are fitting parameters. 
Our definition of the core radius $R$ is the radius 
where the fitting line is recovered to $\Delta_M/C=0.6\Delta_M$. 
Thus, $R$ is given by 
$R=B \ln \sqrt{({C^{1/A}+1})/({C^{1/A}-1})}$.

\begin{figure}[htb]
\includegraphics[width=6cm]{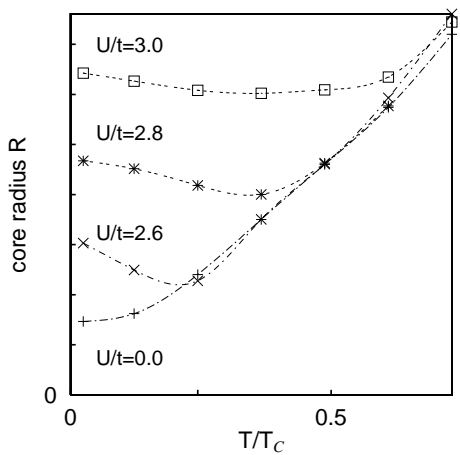}
\caption{\label{fig:radius}
Temperature dependence of the core radius $R$ 
for $U=0$, 2.6, 2.8 and 3.0. 
The departure from the monotonic decreasing curve 
of $U=0.0$ occurs at the corresponding $T_M$.
}
\end{figure}

We show the temperature dependence of the core radius 
in Fig. \ref{fig:radius}. 
When $U=0$, the core radius $R$ monotonically shrinks 
as $T$ lowers, according to the Kramer-Pesch (KP) effect \cite{KP} and 
saturates at lower $T$ because of the quantum limit behavior\cite{HayashiBdG} 
even in the $d$-wave superconductor. 
For finite $U (>U_{\rm cr})$ cases shown in Fig. \ref{fig:radius},  
the deviation from the KP linear curve occurs below the temperature $T_M$ 
where $M_{\rm v}$ appears (see Fig. \ref{fig:UT}(b)).
That is, the shrinkage of $R$ stops at $T_M$, and the core radius increases 
as $T$ lowers. 
This is caused by the growth of the moment induced around the core
which destroys the condensation locally, enlarging $R$.
This feature coincides with the $\mu$SR experiment on 
{\rm La$_{2-x}$Sr$_x$CuO$_4$} by Kadono {\it et al}.,\cite{kadono}
where as $x$ decreases the $R$ stops at higher $T$ and becomes larger. 
If we interpret the doping dependence in terms of the $U$-change, 
the smaller $x$ should correspond to the larger $U$.
This correspondence is natural because 
as going into underdopings the stripe tendency is strengthened.
Tsuchiura {\it et al}.,\cite{tsuchi,tanaka} assert that 
the core radius becomes smaller as approaching to half-filling 
in their t-J model calculation, 
it contradicts $\mu$SR experimental results 
and our extended Hubbard model calculation.

The order parameter $\Delta_d$ vanishes at the vortex core and 
recovers to the bulk value continuously.
The spatial variation of the $\Delta_d$ induces 
the order parameters with other symmetry, such as an extended $s$-wave or 
$p_x \pm{\rm i}p_y$-wave components, around the core.\cite{takigawaLT23} 
Here, we only discuss the $d$-wave triplet pairing component. 
As increasing the repulsive interaction $U$, 
the magnetization arises around the vortex core as mentioned before.
In the range of $U > U_c$, the checkerboard pattern appears on the floor. 
And in $U/t > 3.1$, 
the stripe structure which is modulated by vortices appears.
In these cases, small triplet-$d$-wave component $\Delta_d^{\rm triplet}$ 
appears around the core.\cite{takigawaLT23,ghosal} 
Figure \ref{fig:OP} shows the profile of 
$|\Delta_d^{\rm triplet}({\bf r}_i)|$.
By increasing $U$,
the maximum point in $|\Delta_d^{\rm triplet}|$ 
moves to outside the core 
and the amplitudes become large.

\section{\label{sec:subsec2}Local density of states}

\begin{figure*}[htb]
\includegraphics[width=15cm]{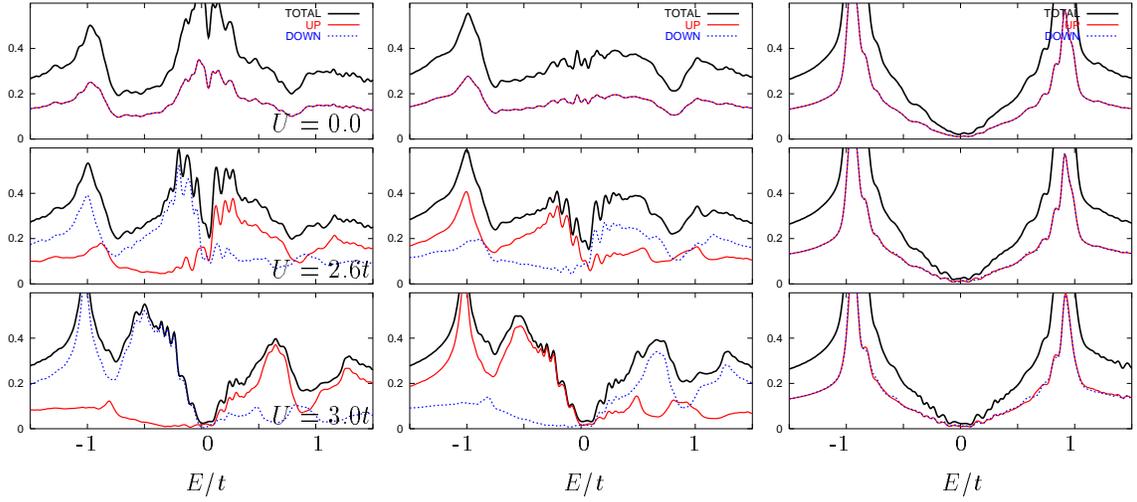}
\caption{\label{fig:LDOS}
The local density of states $N(E,{\bf r}_i)$ for $U/t=0.0,2.6,3.0$ 
at $T=T_c/40$. 
(a) Vortex center site (1,1),  
(b) Nearest neighbor site (1,2) to the vortex center, 
(c) Bulk site (1,12) in the midpoint between next NN vortices.
We also show the up- and down-spin contributions, 
$N_\uparrow(E,{\bf r}_i)$ and $N_\downarrow(E,{\bf r}_i)$. 
The energy is scaled by $t$.
}
\end{figure*}

We study the effect of the induced AFM in the LDOS. 
Figure \ref{fig:LDOS}(a) shows the LDOS at the vortex core site. 
For $U=0$ in the upper panel, we see the ZEP at $E \sim 0$, 
which is a typical feature of the vortex in $d$-wave superconductors.\cite{Franz} 
The small suppression at $E \sim 0$ may come from the induced other 
component, such as an extended $s$-wave, around the vortex 
core.\cite{Himeda} 
However, this effect is small. 
The charge density for each spin is given by 
\begin{eqnarray}&& 
n_{i,\uparrow}=\int{\rm d}E N_\uparrow(E,{\bf r}_i) f(E),   
\nonumber \\ &&
n_{i,\downarrow}=\int{\rm d}E N_\downarrow(E,{\bf r}_i) f(E). 
\label{eq:n-LDOS} 
\end{eqnarray}
When we see the spectrum inside the superconducting gap at $U=0$, 
the weight of $N(E,{\bf r})$ for $E<0$ is smaller 
than that for $E>0$, reflecting the suppression of $n({\bf r}_i)$ 
around the core shown in Fig. \ref{fig:r}(a). 

When the AFM appears at $U>U_{\rm cr}$, 
the ZEP is suppressed around $E=0$. 
With increasing $U$, the suppression becomes eminent, 
and the peaks are shifted to larger $|E|$, 
as shown in the middle and lower panels in Fig. \ref{fig:LDOS}(a). 
When $M({\bf r}_i) \ne 0$, 
$N_\uparrow(E,{\bf r}_i) \ne N_\downarrow(E,{\bf r}_i)$.  
Since $n_{\uparrow,i}< n_{\downarrow,i}$ 
in the case $S_z({\bf r}_i)<0$ at the site of the vortex center, 
$N_\uparrow(E,{\bf r}_i) < N_\downarrow(E,{\bf r}_i)$ for $E<0$ 
and $N_\uparrow(E,{\bf r}_i) > N_\downarrow(E,{\bf r}_i)$ for $E>0$, 
following the relation in Eq. (\ref{eq:n-LDOS}). 
Therefore, split peaks at positive and negative energy, 
respectively, come from the up-spin and down-spin contribution. 
After the AFM appears, the weight of the LDOS for $E>0$ shifts 
to that for $E<0$, reflecting the enhancement of $n({\bf r}_i)$ 
around the core shown in Figs. \ref{fig:r}(c) and \ref{fig:r}(d). 

In Fig. \ref{fig:LDOS}(b), we show the LDOS at the site next to the 
vortex center site. 
There, while the ZEP is smeared at $U=0$, the suppression of the LDOS 
around $E=0$ is seen as in the vortex center site. 
Since this site has up-spin moment, 
split peaks at positive and negative energy, 
respectively, comes from the down-spin and up-spin contribution. 
The LDOS at farthest site in the midpoint between next NN vortices is 
shown in Fig. \ref{fig:LDOS}(c), 
which has spectrum as in the uniform $d$-wave superconductor at zero field. 
By the induced AFM, the low energy state is slightly suppressed.

The spatial structure of the LDOS at $E \sim 0$ is also 
displayed in Fig. \ref{fig:r}. 
The ZEP height at the vortex core becomes low as the moment $M_{\rm v}$ grows. 
In Figs. \ref{fig:r} (c) and \ref{fig:r}(d) 
where the induced moment $M_{\rm v}$ is rather large, 
the peak structure at the core is removed completely and the LDOS 
at the core is small compared with its surroundings, 
giving rise to a ``caldera'' type landscape. 
This is an origin of the so-called 
``empty'' core observed by STM,\cite{renner} as discussed later.

\section{\label{sec:NMR}Nuclear magnetic relaxation time and the resonance line shape}

\begin{figure}[tbh]
\includegraphics[width=8cm]{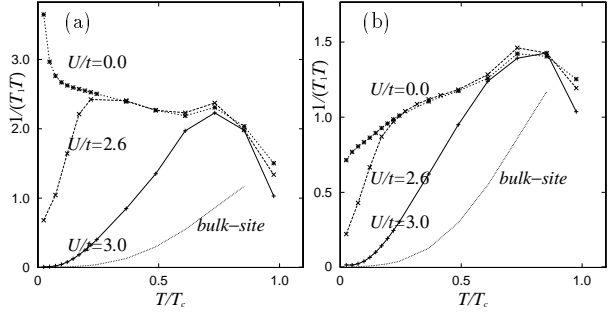}
\caption{\label{fig:T1-T} 
Temperature dependence of $1/(T_1T)$ for $U/t=0$, 2.6 and 3.0.  
(a) Vortex center site (1,1), 
(b) NN-site (1,2) next to the center. 
The $1/(T_1T)$ is normalized by the value at $T_c$. 
The line denoted as ``bulk site'' shows $1/(T_1T)$ at the 
furthest site from the vortex in the midpoint between next NN 
vortices. 
$T_M/T_c \sim$0.25 ($U/t=2.6$) and 0.75 ($U/t=3.0$)
}
\end{figure}

\begin{figure}[bth]
\includegraphics[width=8cm]{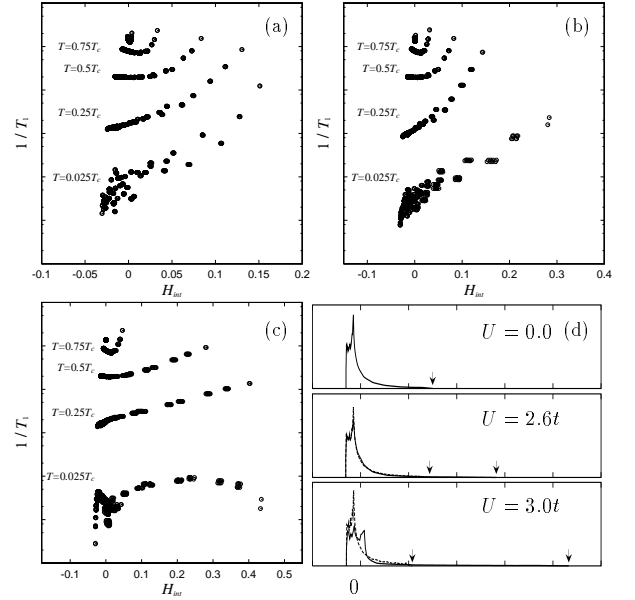}
\caption{\label{fig:NMR}
The distribution of $1/T_1$ as a function of the internal magnetic 
field $H_{\rm int}$ at $T/T_c=0.025$, 0.25, 0.5, 0.75 and 0.95 
in the case $U/t$=0.0 (a), 2.6 (b), and 3.0 (c).
Vertical axis is in log-scale, and horizontal axis is in arbitrary unit.  
The internal field distribution function (d) (resonance line shape of NMR) 
at $T/T_c=0.025$ for $H_{\rm int}({\bf r})$ (solid lines) and 
$H_{\rm int}^{j}({\bf r})$ (dashes lines).
The right side (left side) arrow shows the maximum of $H_{\rm int}$ ($H_{\rm int}^{j}$).
}
\end{figure}

As mentioned in the introduction, 
NMR is a powerful method to simultaneously 
know the field distribution in vortex lattice through the resonance pattern 
(Redfield pattern) and the spatial profile of the zero-energy quasi-particles 
through the relaxation time $T_1$. 
In this section we calculate the site-selective $T_1$ 
based on the solutions obtained in the previous sections 
and examine the outcomes from the ``anomalous'' $d$-wave vortex core.

In Fig. \ref{fig:T1-T}, we plot the $T$-dependence of 
the $1/(T_1T)$. 
Far from the vortex core (line denoted as ``bulk site'' in the figure), 
$T_1$ shows the conventional $T_1^{-1} \sim T^3$ behavior 
of the line node case as in the bulk at zero field. 
As shown in Fig. \ref{fig:T1-T}(a), $1/(T_1T)$ shows peak 
below $T_c$ at the core site. 
And, when $U/t=0$, $1/(T_1T)$ is increased on lowering $T$ due 
the presence of ZEP. 
When $M_{\rm v}$ appears at $T_M (<T_c)$, 
as shown in Fig. \ref{fig:UT}(b), 
$1/(T_1T)$ behavior deviates from the line of the $U=0$ case at $T_M$. 
Below $T_M$, $1/(T_1T)$ is suppressed with lowering $T$, 
reflecting the decrease of the zero-energy DOS by the splitting of ZEP 
in Fig. \ref{fig:LDOS}(a). 
Therefore, we see that, on lowering $T$, 
$1/(T_1T)$ increases at $T>T_M$ and decreases at $T<T_M$ 
at the vortex core site, 
which is consistent with the observed $1/(T_1T)$ behavior.\cite{kakuyanagi2} 
At NN-site (Fig. \ref{fig:T1-T}(b)),  
$1/(T_1T)$ is also suppressed at $T<T_M$ by losing DOS for $E \sim 0$.  

Next, we study the internal field dependence of $1/(T_1T)$. 
The $1/(T_1T)$ can be distinguished spatially 
by its resonant frequency in NMR experiment.
The frequency depends on the internal magnetic field $H_{\rm int}({\bf r})$, 
which consists of a contribution from 
the screening current in the vortex lattice 
and the direct contribution from the AFM-moment.
At first, we evaluate $H_{\rm int}^{j}({\bf r})$ 
from the contribution via the screening current 
distribution through the Maxwell equation: 
\begin{equation}
\nabla \times {\bf H}= \frac{4\pi}{c} {\bf j}({\bf r}).
\end{equation}
The current is calculated as 
\begin{eqnarray} 
j_{\hat{e}}({\bf r}_i)
&=& 
2 |e| c {\rm Im}\{ {\tilde t}_{i+\hat{e},i} 
\sum_\sigma \langle a^\dagger_{i+{\hat e},\sigma}
a_{i,\sigma} \rangle \} \\
&=& 
2 |e| c {\rm Im}\{ {\tilde t}_{i+{\hat e},i} 
\sum_\alpha [ u^\ast_\alpha({\bf r}_{i+{\hat e}})u_\alpha({\bf r}_{i}) 
f(E_\alpha) \nonumber \\ &&
+ v_\alpha({\bf r}_{i+{\hat e}})v^\ast_\alpha({\bf r}_{i}) 
(1-f(E_\alpha)) ] \}
\label{eq:current}
\end{eqnarray}
for the $\hat{e}$-direction bond ($\hat{e}=\pm\hat{x}$, $\pm\hat{y}$) 
at the site ${\bf r}_i$. 
The screening current $j_{\hat{e}}({\bf r}_i)$ 
circles around each vortex core.
When $M_{\rm v} \neq 0$, $j_{\hat{e}}({\bf r}_i)$ exhibits the 
staggered-like fluctuation current.\cite{kishine} 
It does not appear when $M_{\rm v}=0$ even if $U/t>0$.

Figure \ref{fig:NMR} shows $1/T_1$ 
as a function of the internal field $H_{\rm int}$ 
for some temperatures. 
Here, we take account of the contribution from the AFM-moment. 
The internal field is given by
\begin{eqnarray}
H_{\rm int}({\bf r})&=& 
\sqrt{(H_{\rm int}^{j}({\bf r})+H_0)^2 + m^2 M^2({\bf r})}-H_0
 \nonumber \\
&\simeq& H_{\rm int}^{j}({\bf r}) + \frac{m^2}{2H_0}M^2({\bf r})
\label{eq:H}
\end{eqnarray}
for $H_{\rm int}^{j}({\bf r}),|mM({\bf r})| \ll H_0$.
$H_0$ is the external field.
$H_{\rm int}^{j}({\bf r})$ is a contribution of 
the screening current $j_{\hat{e}}$ in the vortex lattice.
The $mM({\bf r})$ with a numerical factor $m$ 
is the direct contribution from AFM moment.
Here, we assume that the magnetic moments are parallel to 
the {\rm CuO}-layer as in the AF-state at the half-filling. 
And we choose $m$ to reproduce the observed 
line shape of NMR.\cite{kakuyanagip}

The internal field distribution function, corresponding to 
the resonance line shape of NMR, 
is shown for lowest temperature case in Figs. \ref{fig:NMR}(a)-(c).
The horizontal width of $1/T_1$-distribution corresponds to 
the width of the resonance line shape for higher temperature data.  
As increasing temperature, 
the width of the resonance line shape becomes narrow (Fig. \ref{fig:NMR}(a))
since the internal field narrowly distributes in the vortex lattice. 
The width depends on the radius of the core. 
As the core radius becomes larger, the maximum point of 
the screening current becomes far from the vortex center, and 
the peak of the internal field at the vortex center becomes broad. 
Thus, the width of the resonance line shape becomes narrow. 

This relation between the width of the resonant line shape 
and the vortex core radius appears also in the $U$-dependence, 
if we neglect the contribution from $M({\bf r})$. 
We compare the field distribution function of $H_{\rm int}^{j}({\bf r})$ 
at $U/t=0$, $2.6$ and $3.0$, which are shown by dashed lines in Fig. \ref{fig:NMR}(d). 
In this figure left side arrows show the maximum of $H_{\rm int}^{j}({\bf r})$.
In the case $U/t=2.6$ and $3.0$ where 
the moment appears around the core at low temperatures, 
the core radius $R$ is larger, 
and the width of the line shape becomes narrower at low temperature.

The realistic line shape observed in NMR 
includes the contribution of the AFM moment.
The solid lines in Fig. \ref{fig:NMR}(d) 
show the field distribution function of $H_{\rm int}({\bf r})$ 
given by Eq. \ref{eq:H}.
In this figure, right side arrows show the maximum of $H_{\rm int}({\bf r})$. 
In the case $U/t=2.6$ and $3.0$, 
by the AFM effect, the resonance line shape becomes broader and have 
a long tail toward the higher-$H_{\rm int}$ side.
After the magnetization appears, the ZEP in the LDOS splits and $1/T_1$ 
becomes small. When we see the $1/T_1$-behavior as a function of $H_{\rm int}$, 
as in Fig. \ref{fig:NMR}, $1/T_1$ shows the same behavior as the $U=0$ case 
for $T>T_M$. 
At $T<T_M$, 
the effective internal field is shifted toward higher-$H_{\rm int}$ side, 
and $1/T_1$ is decreased.
These AFM contributions are eminent at higher-$H_{\rm int}$ region, 
where signals come from the spatial region around the vortex core.
In Fig. \ref{fig:NMR}(c), as $H_{\rm int}$ increases, $1/T_1$ decreases near the vortex 
core region (high $H_{\rm int}$ region) at low temperature, 
reflecting the special feature of the spatial structure, 
namely ``caldera'' structure of $N(E=0,{\bf r})$ shown in 
Fig. \ref{fig:r}(c). 
This special feature is observed by Kakuyanagi {\it et al}.\cite{kakuyanagip}
The spatial structure of this 1/$T_1$({\bf r})-behavior 
appears at lowest temperature 
when $U$ is strong (near the stripe state). 
To reproduce the suppression of $1/T_1$ at the vortex core region, 
the choice of the Fermi surface shape (i.e. $t'$ and $t''$) is important.
For example, we cannot obtain "caldera" type-LDOS structure when $t'=0$.

\section{\label{sec:con}Conclusion and discussion}

In this paper, based on the extended Hubbard Hamiltonian, 
we have studied the vortex structure 
when the incommensurate AFM is induced around the vortex core 
by the competition between AFM and 
superconductivity due to the on-site repulsive interaction $U$.  
We calculate the spatial distribution of the $d$-wave 
superconducting order parameter, the magnetization, charge density, 
the local density of states and the nuclear spin relaxation rate $T_1$, 
and clarify the $U$- and temperature-dependences.  

As increasing $U$, 
the AFM with weak checkerboard modulation appears around 
the vortex core at $U > U_{\rm cr}$, where 
the AFM-moment appears at $T_M (<T_c)$. 
At further large $U$, the checkerboard modulation changes to 
the stripe modulation, where  $T_M > T_c$.  
This transitions is clearly seen by analyzing the Fourier component 
corresponding the ordering vector of the incommensurate modulation. 

By the induced AFM around the vortex core, 
the ZEP at the vortex core at $U=0$ splits, 
and the zero-energy density of states is suppressed. 
Reflecting this LDOS structure, $1/T_1$ is suppressed at the vortex core 
in low temperature range. 
That is, on lowering $T$, $1/T_1$ is enhanced at $T>T_M$ and 
decreased at $T<T_M$. 
These $T_1$-behaviors are consistent with the 
observation.\cite{kakuyanagi2}
We also discuss the AFM effect on the internal field distribution, 
relating to the resonance line shape in NMR.

The spectrum structure of the LDOS is related to the 
spin and charge structure, such as the suppression (enhancement) 
of the charge density around the vortex core when the AFM 
is absent (present). 
The difference of the LDOS between up-spin and down-spin 
is essentially important especially to discuss the transport phenomena.

We also show that the vortex core radius becomes large 
as the AFM appears around the core. 
This behavior explains the doping dependence 
in recent $\mu$SR experiment. 
In this way, the induced AFM contribution plays an important role 
in the study of the various physical properties in the vortex state 
of high $T_c$-cuprate.

\begin{acknowledgments}
We thank 
K. Kakuyanagi, K. Kumagai, Y. Matsuda, 
K. Ishida and V.F. Mitrovi\'c on their NMR 
for useful discussions.
We thank 
R. Kadono on their $\mu$SR 
for useful discussions.
We thank 
Y. Tanaka and M. Ogata, 
for useful discussions.
\end{acknowledgments}


\end{document}